\documentclass[12pt]{iopart}
\usepackage{iopams}
\usepackage{graphicx}
\input amssym.def \input amssym

\begin{document}

\title[]{Pauli graphs, Riemann hypothesis, Goldbach pairs}

\author{Michel Planat$^{1}$, Fabio Anselmi$^{1}$
and Patrick Sol\'e$^{2}$
}

\address{ $^1$ Institut FEMTO-ST, CNRS, 32 Avenue de l'Observatoire, F-25044 Besan\c con, France.}

\address{$^2$ Telecom ParisTech, 46 rue Barrault, 75634 Paris Cedex 13, France.}

\begin{abstract}

Let consider the Pauli group $\mathcal{P}_q=\left\langle X,Z\right\rangle$ with unitary quantum generators $X$ (shift) and $Z$ (clock) acting on the vectors of the $q$-dimensional Hilbert space via $X\left|s\right \rangle=\left|s+1\right \rangle$ and $Z\left|s\right \rangle=\omega^s \left|s\right \rangle$, with $\omega=\exp(2i\pi/q)$. It has been found that the number of maximal mutually commuting sets within $\mathcal{P}_q$ is controlled by the Dedekind psi function $\psi(q)=q \prod_{p|q}(1+\frac{1}{p})$ (with $p$ a prime) \cite{Planat2011} and that there exists a specific inequality $\frac{\psi (q)}{q}>e^{\gamma}\log \log q$, involving the Euler constant $\gamma \sim 0.577$, that is only satisfied at specific low dimensions $q \in \mathcal {A}=\{2,3,4,5,6,8,10,12,18,30\}$. The set $\mathcal{A}$ is closely related to the set $\mathcal{A} \cup \{1,24\}$ of integers that are totally Goldbach, i.e. that consist of all primes $p<n-1$ with $p$ not dividing $n$ and such that $n-p$ is prime \cite{Brouwers2004}. In the extreme high dimensional case, at primorial numbers $N_r$, it is known that the inequality $\frac{\psi(N_r)}{N_r \log \log N_r} \gtrapprox \frac{e^{\gamma}}{\zeta(2)}$ (for every $r>2$) is equivalent to Riemann  hypothesis. Introducing the Hardy-Littlewood function $R(q)=2 C_2 \prod_{p|n}\frac{p-1}{p-2}$ (with $C_2 \sim 0.660$ the twin prime constant), that is used for estimating the number $g(q) \sim R(q) \frac{q}{\ln^2 q }$ of Goldbach pairs, one shows that the new inequality $\frac{R(N_r)}{\log \log N_r} \gtrapprox e^{\gamma}$  is also equivalent to Riemann hypothesis. In this paper, these number theoretical properties are discusssed in the context of the qudit commutation structure.

\end{abstract}

\pacs {02.10.De, 02.10.0x, 03.65.Fd, 03.67.Lx}

~~~~~~~~~~~~\footnotesize{ Mathematics Subject Classification: 11M26, 11P32, 05C10, 81P68, 11A25}

\normalsize

\noindent

\section{Introduction}

We propose new connections between the Pauli graphs \cite{PlanatQIC,Planat2011}, that encode the commutation relations of qudit observables, and prime number theory. We already emphasized that the Dedekind psi function $\psi(q)=q \prod_{p|q}(1+\frac{1}{p})$ (with $p$ a prime) is used to count the number of maximal commuting sets of the qudits  \cite{Planat2011} and meets the Riemann hypothesis (RH) at primorial numbers $q\equiv N_r=2\cdots p_r$ \cite{PlanatSole2010}. Similarly, there exist striking connections between $\psi(q)$ and the Hardy-Littlewood function $g(q)=R(q)\frac{q}{\log^2(q)}$ for the Goldbach distribution of prime pairs [see Sec. 3 for the definition of $R(q)$]. In particular, we observe that $\psi(q)$ meets the so-called totally Goldbach numbers at small $q$'s and that $R(q)<\zeta(2)\frac{\psi(q)}{q}$ also meets RH at primorial numbers.

The Euler constant $\gamma=\lim_{n \rightarrow \infty}(\sum_{k=1}^n \frac{1}{k} -\log n)\sim 0.57721$, through the Mertens formula $e^\gamma=\lim_{n \rightarrow \infty} \frac{1}{\log n}\prod_{p\le n}(1-\frac{1}{p})^{-1}$, is an important ingredient of all the inequalities involved in this correspondence.

In the rest of this section, we report on the number theoretical {\it coincidence} between $\psi(q)$ and the totally Goldbach numbers at small $q$'s, as well on the already known theorem connecting $\psi(q)$ and RH at primorial numbers $N_r$. In Section 2, we explore in detail this coincidence by refering to the qudit Pauli graphs. Then, in Section 3, we establish the connection between $R(q)$ and RH at primorial numbers. In the discussion, the concept of a Goldbach defect for encompassing the statements at low and high $q$'s is proposed.

\subsection*{A number theoretical coincidence}


Let us start the exposition of our ideas with a few theorems and definitions.

Goldbach's conjecture, formulated in 1742, is that every even integer greater than $2$ is the sum of two primes. To date it has been checked for $q$ up to $2.10^{18}$ \cite{Oliveira}. A pair ($p_1,p_2$) of primes such that the even integer $n=p_1+p_2$ is called a Golbach partition.

{\bf Definition 1:} A positive integer $n$ is {\it totally Goldbach} if for all primes $p<n-1$, with $p$ not dividing $n$ (except when $p=n-p$) we have that $n-p$ is prime \cite{Brouwers2004}.

{\bf Theorem 1:} The inequality $\frac{\psi (q)}{q}>e^{\gamma}\log \log q$ is only satisfied at a totally Goldbach number $q \in \mathcal {A}=\{2,3,4,5,6,8,10,12,18,30\}$. The only totally Goldbach numbers not satisfying the inequality are $q=1$ and $q=24$.

{\bf Proof: } The proof of theorem 1 follows from a combination of results in \cite{Brouwers2004} and \cite{PlanatSole2010} (for a more general setting, see \cite{PlanatSole2011}).

{\bf Definition 2:} A positive integer $n$ is {\it almost totally Goldbach} of index $r$ if for all primes $p<n-1$, with $p$ not dividing $n$ (except when $p=n-p$) we have that $n-p$ is prime with $r$ exceptions.

\begin{table}[ht]
\begin{center}
\small
\begin{tabular}{|r|r|r|}
\hline
 index $r$ & set &  almost totally Goldbach mumbers \\
\hline
$0$ & $\mathcal{A}_0$&   $\{1,2,3,4,5,6,8,10,12,18,24,30\}$\\
$1$ & $\mathcal{A}_1$& $\mathcal{A}_0\cup\{7,9,14,16,20,36,42,60\}$\\
$2$ & $\mathcal{A}_2$& $\mathcal{A}_1\cup\{15,22,48,90\}$\\
$3$ & $\mathcal{A}_3$& $\mathcal{A}_2\cup\{13,26,28,34,54,66,84,120\}$\\
$4$ & $\mathcal{A}_4$& $\mathcal{A}_3\cup\{11,21,40,78,210\}$\\
$5$ & $\mathcal{A}_5$& $\mathcal{A}_4\cup\{19,32,44,50,72\}$\\
$6$ & $\mathcal{A}_6$& $\mathcal{A}_5\cup\{17,25,46,70,102,114\}$\\
$7$ & $\mathcal{A}_7$& $\mathcal{A}_6\cup\{33,38,52,64,126,150\}$\\
$8$ & $\mathcal{A}_8$& $\mathcal{A}_7\cup\{23,27,31,39,56,58,96\}$\\
$9$ & $\mathcal{A}_9$& $\mathcal{A}_8\cup\{29,35,76,108,168,180\}$\\
$10$ & $\mathcal{A}_{10}$& $\mathcal{A}_9\cup\{45,74,132,144\}$\\

\hline
\end{tabular}
\label{table1}
\caption{Almost totally Goldbach numbers of index $r\le 10$.}
\end{center}
\end{table}

Let $g(n)$ be the number of ways of representing the integer $n$ as the sum of two primes. The maximum value of $g(n)$ is indeed less or equal than the number of primes $\frac{n}{2} \le p \le n-1$. Values of $n$ such that $g(n)$ reaches its maximum are in the set $\mathcal{B}=\mathcal{A}_0\cup \{7,14,16,36,42,48,60,90,210\}$ \cite{Deshou}, where $\mathcal{A}_0$ is the set of totally Goldbach numbers. It is not surprising that numbers in $\mathcal{B}$ that are not totally Goldbach are almost totally Goldbach with a small index $r$ (as shown in Table 1). 
The first five and the integer $60$ have index 1, while $48$ and $90$ have index $2$ and $210$ has index $4$. For a prime number $p>3$, the index $r(p)$ is the Sloane's sequence $A062302$.

This number theoretical coincidence is further explored in Sec. 2 in the context of the qudit commutation structure.

\subsection*{Number theoretical inequalities at primorial numbers}

Let us start with 

{\bf Theorem 2:} Let $N_r=2 \cdots p_r$ be the primorial of order $r$. The statement $\frac{\psi(N_r)}{N_r \log \log N_r}>\frac{e^\gamma}{\zeta(2)}$ for every $r>2$ is equivalent to the Riemann hypothesis.

{\bf Proof:} Theorem 2 is proved in \cite{PlanatSole2010} and in a more general form in \cite{PlanatSole2011}.  

In Sec. 3, we establish further statements relating $\psi(q)$ and the Hardy-Littlewood function $R(q)$ about the number of Goldbach pairs. An important result is conjecture 1 that is found to be equivalent to RH by theorem 5.

\section{The number theoretical coincidence for qudits}

In this section, we recall the definition of a $q$-level system, or qudit, and the number theoretical structure, involving the Dedekind psi function $\psi(q)$, of maximal sets of pairwise commuting operators. Calculations about the inequality of theorem 1, at the number $q$, and the symmetries of the qudit system in the corresponding dimension $q$ are given in two separate tables 2 and 3, in order to display the distinctive features of both types of systems. In table 2, one selects $q \in \mathcal{A}_0$, a totally Goldbach number, and in table 3 one selects $q \notin \mathcal{A}_0$, with $q \le 36$.

\subsection*{The single qudit commutation structure}

{\bf Definition 3:} In a loose sense, a qudit is defined as the group $\mathcal{P}_q=\left\langle X,Z\right\rangle$ with unitary quantum generators
\begin{equation}
X=\left(\begin{array}{ccccc} 0 &0 &\ldots &0& 1 \\1 & 0  &\ldots & 0&0 \\. & . & \ldots &.& . \\. & . & \ldots &.& . \\0& 0 &\ldots &1 & 0\\ \end{array}\right),~~ Z= \mbox{diag}(1,\omega,\omega^2,\ldots,\omega^{q-1}),
\label{Paulis}
\end{equation}
with $\omega=\exp(\frac{2i\pi}{q})$. The Weyl pair $(X,Z)$ satisfies $ZX-\omega XZ=0$, and each element of $\mathcal{P}_q$ may be uniquely written as $\omega^aX^bY^c$, with $a,b,c$ in the modular ring $\mathbb{Z}_q$. One has $|\mathcal{P}_q|=q^3$, $\mathcal{P}_q'=Z(\mathcal{P}_q)$ (the derived subgroup equals the center of the group, and is of order $q$). One is interested in the maximal mutually commuting sets within the central quotient $\tilde{\mathcal{P}}_q=\mathcal{P}_q/Z(\mathcal{P}_q)$, with the identity $I_q$ removed.

{\bf Theorem 3:} Let $\mathfrak{S}$ be the set of maximal mutually commuting sets $\mathfrak{M}$ in $\tilde{\mathcal{P}}_q$. The cardinality of $\mathfrak{M}$ is $|\mathfrak{M}|=q-1$ and that of $\mathfrak{S}$ is $|\mathfrak{S}|=\sigma (q)$, where $\sigma(q)$ is the sum of divisor function.

{\bf Proof:} The proof of theorem 3 is given in \cite{Albouy2009} (for the relation to $\sigma(q)$, see \cite{Planat2011}).

Hint: Using the Weyl pair property one writes the group theoretical commutator as $\left[X,Z\right]=XZX^{-1}Z^{-1}=\omega^{-1} I_q$ so that one gets the expression  
$$\left[\omega^aX^bZ^c,\omega^{a'}X^{b'}Z^{c'}\right]=\omega^{cb'-c'b}I_q,$$
meaning that two elements of $\mathcal{P}_q$ commute if only if the determinant $\Delta=\mbox{det}\left(\begin{array}{cc} b' &b \\c'& c \\ \end{array}\right)$ vanishes.
Two vectors such that their symplectic inner product $\left[(b',c').(b,c)\right] =\Delta=b'c-bc'$ vanishes are called perpendicular. Thus, one can transfer the study of commutation relations within the group $\mathcal{P}_q$ to the study of perpendicularity of vectors in the ring $\mathbb{Z}_q^2$ \cite{Havlicek2008} with the isomorphism 
\begin{equation}
(\mathcal{P}_q/Z(\mathcal{P}_q),\times) \cong (\mathbb{Z}_q^2,+)
\label{iso}
\end{equation}
between the central quotient of $\mathcal{P}_q$ and the algebra of vectors in the $\mathbb{Z}_q$-modular lattice $\mathbb{Z}_q^2$, endowed with the symplectic inner product \lq\lq .".

Then, one defines a {\it isotropic line} as a set of $q$ points on the lattice $\mathbb{Z}_q^2$ such that the symplectic product of any two of them is $0 (\mbox{mod}~ q)$. To such an isotropic line corresponds a maximal commuting set in  $\mathcal{P}_q/Z(\mathcal{P}_q)$.

Taking the prime power decomposition of the Hilbert space dimension as $q=\prod_i p_i^{s_i}$, it is shown in (18) of \cite{Albouy2009} that the number of isotropic lines of the lattice $\mathbb{Z}_q^2$ reads 
\begin{equation}
\prod_i \frac{p_i^{s_i+1}-1}{p_i-1}\equiv \sigma (q).
\label{divisor1}
\end{equation}

{\bf Theorem 4:} Let $\mathfrak{S}$ be the set of maximal mutually commuting sets $\mathfrak{M}$ in $\tilde{\mathcal{P}}_q$ and $\mathfrak{D}\subset \mathfrak{S}$ the subset of $\mathfrak{S}$ such that for any  $d \in \mathfrak{D}$ and  $s \in \mathfrak{S}$ the intersection of $d$ and $s$ is non empty, i.e. $d \cap s \ne \emptyset$. The cardinality of $\mathfrak{S}-\mathfrak{D}$ is the Dedekind psi function $\psi(q)$.

{\bf Proof:} 
As shown in \cite{Albouy2009}, a isotropic line of $\mathbb{Z}_q^2$ corresponds to a {\it Lagrangian submodule}, i.e. a maximal module such that the perpendicular module $M^{\perp}=M$. Let us now specialize to Lagrangian submodules that are {\it free cyclic submodules}
$$\mathbb{Z}_q(b,c)=\left\{(ub,uc)|u \in \mathbb{Z}_q\right\},$$
for which the application $u \rightarrow (ub,uc)$ is injective. Not all Lagrangian submodules are free cyclic submodules. A point $x=(b,c)$ such that $\mathbb{Z}_q(b,c)$ is free is called an {\it admissible vector}, and the set of free cyclic submodules defined by the admissible vectors is called the projective line
$$\mathbb{P}_1(\mathbb{Z}_q)=\left\{\mathbb{Z}_q(b,c)|(b,c) ~\mbox{is}~\mbox{admissible}\right\}.$$
Following theorem 5 in \cite{Havlicek2008}, the number of points of the projective line is
\begin{equation}
|\mathbb{P}_1(\mathbb{Z}_q)|=\prod_i (p_i^{s_i}+p_i^{s_i-1})\equiv \psi(q),
\label{proj1card}
\end{equation}
where $\psi(q)$ is the Dedekind psi function (see also \cite{Vourdas2010}).

Then, using the isomorphism relation (\ref{iso}) between the isotropic lines of $\mathbb{Z}_q^2$ and the maximal commuting sets of $\mathfrak{S}$, it is clear that a free cyclic submodule is isomorphic to an element of the set $\mathfrak{S}-\mathfrak{D}$, and that the projective line is in bijective correspondance with $\mathfrak{S}-\mathfrak{D}$. This completes the proof.

\subsection*{Pauli graphs over a single qudit system}

{\bf Definition 4:} Let $\mathcal{G}_q$ be the Pauli graph constructed by taking the observables as vertices and an edge joining two mutually commuting observables (with the identity $I_q$ removed). Indeed,  $|\mathcal{G}_q|=q^2-1$.
One further defines a point/line incidence geometry with observables as points and the maximum cliques as lines. One characterizes this geometry by creating a dual graph $\mathcal{G}_q^\star$ such that the vertices are the cliques (there are $\sigma(q)$ of them) and an edge joins two non-intersecting maximum cliques. The connected component of $\mathcal{G}_q^\star$ corresponds to the graph of the projective line $\mathbb{P}_1(\mathbb{Z}_q)$ [of size $\psi(q)$].

\begin{table}[ht]
\begin{center}
\small
\begin{tabular}{|r|r|r|r|r|r|}
\hline
 $q$ & $\#$tg &  $\#$mc & $\epsilon(q)$ & aut($\mathcal{G}_q^*$)& graph \& subgraphs\\
\hline
$2$ & $0$ &   $3$ &$2.15$ & $S_3$ & $K_3$\\
$3$ & $0$ &   $4$ &$1.16$ & $S_4$ & $K_4$\\
$4$ & $1$ &   $6+1$ &$0.92$ & $G_{48}=\mathbb{Z}_2 \times S_4$ & $K_{2,2,2}$\\
$5$ & $2$ &   $6$ &$0.35$ & $S_6$ & $K_6$\\
$6$ & $1$ &   $12$ &$0.96$ & $G_{144}=A_4 \times D_6$ or $S_4 \times S_3$ & $\mathcal{G}_6^* \equiv \mbox{Gr}_{4,3}$, $3 \times K_4$\\
$8$ & $2$ &   $12+3$ &$0.20$ & $\mathbb{Z}_2^6 \rtimes(\mathbb{Z}_3^3\rtimes G_{48})$ & $\mathcal{G}_8^*$, $3 \times C_4$\\
$10$ & $3$ &   $18$ &$0.31$ & $A_6 \rtimes D_6$ or $ S_6 \times S_3$ & $\mathcal{G}_{10}^* \equiv \mbox{Gr}_{6,3}$, $3 \times K_6$\\
$12$ & $2$ &   $24+4$ &$0.38$ & $\mathbb{Z}_2^{12} \times G_{144}$ & $3 \times \mbox{Cu}$, $4 \times \mathcal{G}_4^*$, $6 \times K_4$\\
$18$ & $4$ &   $36+3$ &$0.11$ & $\mathbb{Z}_3^{12}\rtimes(\mathbb{Z}_2^{12} \times G_{144})$ & $3 \times \mathcal{G}_9^*$, $4 \times \mbox{Gr}_{3,3}$, $12 \times K_3$\\ 
$24$ & $6$ &   $48+12$ &$-0.059$ & $(\mathbb{Z}_2^{24}\rtimes\mathbb{Z}_3^{12})\rtimes(\mathbb{Z}_2^{12} \times G_{144})$ & $3 \times \mbox{BP}_6$, $4 \times \mathcal{G}_8^*$, \\
&&&&& $6 \times \mbox{Cu}$ ,$12 \times \{ K_4, K_{2,2}\}$\\
$30$ & $6$ &   $72$ &$0.22$ & $S_3 \times S_4 \times S_6$ & $4 \times \mathcal{G}_{10}^*$, $6 \times \mathcal{G}_{6}^*$, $12 \times K_6$\\
\hline
\end{tabular}
\label{table2}
\caption{Data about the structure maximal mutually commuting sets for qudits when $q \in \mathcal{A}_0$ (the set of totally Goldbach numbers). The symbols $\#$tg and $\#$mc denote the number of totally Goldbach pairs and the number of maximal mutually commuting sets, respectively. In column 4, one computes the difference $\epsilon (q)=\frac{\psi (q)}{q}-e^{\gamma}\log \log q$ and in column 5 the automorphism group of the corresponding projective line. The graphs or the non trivial subgraphs that may identified at the single or multiple point intersection set are displayed in colum 6. The notations $K_n$, $K_{n_1,\cdots,n_l}$, $C_n$, $\mbox{Gr}_{p,q}$ and $\mbox{Cu}$ are for the complete graph with $n$ vertices, a $l$-partite graph, the polygon graph with $n$ vertices, the grid graph $p\times q$ and the cube graph respectively, respectively. The notation $\mbox{BP}_6$ means a bipartite graph of spectrum $\{0^8,(-6)^1,6^1,(-2)^3,2^3\}$. Here, the symbol $\times$ corresponds to non-intersecting copies of the relevant subgraph. }
\end{center}
\end{table}

\begin{table}[ht]
\begin{center}
\small
\begin{tabular}{|r|r|r|r|r|r|}
\hline
 $q$ & $(r,\#$atg) &  $\#$mc & $\epsilon(q)$ & aut($\mathcal{G}_q^*$)& graph \& subgraphs\\
\hline
$7$ & $(1,2)$ &   $8$ &$-0.043 $& $S_8$ & $K_8$\\
$9$ & $(1,2)$ &   $12+1$ &$-0.069$ & $(\mathbb{Z}_3^4 \rtimes \mathbb{Z}_2^3)\rtimes G_{48}$ & $\mathcal{G}_9^*$, $4 \times K_3$\\
$11$ & $(4,0)$ &   $12$ &$-0.47 $& $S_{12}$ & $K_{12}$\\
$13$ & $(3,2)$ &   $14$ &$-0.60$ & $S_{14}$ & $K_{14}$\\
$14$ & $(1,3)$ &   $24$ &$-0.014 $& $A_8\rtimes D_6$ or $S_8 \times S_3$ &  $\mathcal{G}_{14}^*$ ,$3 \times K_8$\\
$15$ & $(2,2)$ &   $24$ &$-0.17$ & $A_6 \rtimes G_{48}$ or $S_6 \times S_4$ &$\mathcal{G}_{15}^*$ ,$4 \times K_6$\\
$16$ & $(1,4)$ &   $24+7$ &$-0.32$ & $A_8^2 \rtimes G_{48}$ & $3 \times K_{4,4}$, $6 \times C_4$\\
$17$ & $(6,0)$ &   $18$ &$-0.80$ & $S_{18}$ & $K_{18}$\\
$19$ & $(5,2)$ &   $20$ &$-0.87$ & $S_{20}$ & $K_{20}$\\
$20$ & $(1,4)$ &   $36+4$ &$-0.15$ & $\mathbb{Z}_2^{18} \rtimes (S_6^2 \times S_3$ & $3 \times \mbox{BP}_5$, $6 \times \{K_6,\mathcal{G}_4^*\}$\\
$21$ & $(4,2)$ &   $32$ &$-0.46$ & $A_8 \rtimes G_{48}$ or $S_8 \times S_4$ & $\mathcal{G}_{21}^*$, $4 \times K_8$\\
$22$ & $(2,5)$ &   $36$ & $-0.37$& $A_{12}\rtimes D_6$ or $S_{12} \times S_3$ & $\mathcal{G}_{22}^*$\\
$23$ & $(8,0)$ &   $24$ & $-0.99$& $S_{24}$ & $K_{24}$\\
$25$ & $(5,2)$ &   $30+1$ &$-0.88$ & $A_5^6 \rtimes(\mathbb{Z}_2^6 \rtimes S_6)$ & $6 \times K_5$\\
$26$ & $(3,5)$ &   $42$ & $-0.49$& $A_{14}\rtimes D_6$ or $S_{14} \times S_3$ & $\mathcal{G}_{26}^*$, $3 \times K_{14}$\\
$27$ & $(8,0)$ &   $36+4$ &$ -0.79$& $A_9^4 \rtimes(\mathbb{Z}_2^4\rtimes S_4)$ & $4 \times K_{3,3,3}$\\
$28$ & $(3,4)$ &   $48+8$ & $-0.42$& $\mathbb{Z}_2^{24}\rtimes(S_8\times S_3 )$ &  $3 \times \mbox{BP}_7$, $6 \times K_8$, $8 \times \mathcal{G}_4^*$\\
$29$ & $(9,0)$ &   $30$   & $-1.13$& $S_{30}$ & $K_{30}$ \\

$31$ & $(8,2)$ &   $32$ & $-1.16$& $S_{32}$ & $K_{32}$\\
$32$ & $(5,4)$ &   $48+15$ &$-0.71$ & $A_{16}^3 \rtimes G_{48}$ & $3 \times K_{8,8}$, $6 \times K_{4,4}$, $12 \times C_4$\\

$33$ & $(7,2)$ &   $48$ &$-0.77$ & $A_{12}\rtimes G_{48}$ or $S_{12} \times S_4$ &$\mathcal{G}_{33}^*$, $4 \times K_{12}$\\
$34$ & $(3,7)$ &   $54$ &$-0.65$ & $A_{18}\rtimes D_6$ or $S_{18} \times S_3$ & $\mathcal{G}_{34}^*$,  $3 \times K_{18}$\\
$35$ & $(9,0)$ &   $48$ &$-0.89$ & $A_8 \rtimes G_{48}$ or $S_8 \times S_6$  & $\mathcal{G}_{35}^*$,  $6 \times K_{8}$\\
$36$ & $(1,8)$ &   $72+19$ &$-0.27$ & $A_6^{12}\rtimes (\mathbb{Z}_2^{10}\rtimes S_4^2)$ & $4 \times \mbox{SG}$, $3\times \mbox{BP}_9$, $12 \times \{  C_6, \mathcal{G}_4^*\}$\\

\hline
\end{tabular}
\label{table3}
\caption{Data about the structure of maximal mutually commuting sets for qudits when $q \notin \mathcal{A}_0$. The two symbols r and $\#$atg denotes the index of a almost totally Goldbach pair (see definition 2) and the number of them.  The symbol $\#$mc   denotes the number of maximal mutually commuting sets. In column 4, one computes the difference $\epsilon (q)=\frac{\psi (q)}{q}-e^{\gamma}\log \log q$ and in column 5 the automorphism group of the corresponding projective line. The graphs or the non trivial subgraphs that may identified at the single or multiple point intersection set are displayed in colum 6. Notations are the same than for table 2. The notation $\mbox{BP}_r$ means the bipartite graph of prime index $r$ and spectrum $\{-r^1,r^1,(-1)^r,1^r\}$; the graph $\mbox{BP}_3$ is the cube graph $\mbox{Cu}$. The notation $\mbox{BP}_9$ means a bipartite graph of spectrum $\{0^{16},(-9)^1,9^1,(-3)^3,3^3\}$. The spectrum of the subgraph $\mbox{SG}$ found for $q=36$ is $\{8^1,0^9,(4)^4,2^4\}$.}
\end{center}
\end{table}

{\bf Definition 5:} Let $\mathcal{G}$ be a graph. The set of graph eigenvalues $\lambda_i$ (with multiplicities $n_i$), also called the graph spectrum of $\mathcal{G}$, is denoted $\mbox{spec}(\mathcal{G})=\{\lambda_1^{n_1},\lambda_2^{n_2},\ldots\}$. A graph whose spectrum consists entirely of integers is called an integral graph. 

\subsection*{Results}

Main properties of graphs $\mathcal{G}_q$ and $\mathcal{G}_q^*$ are displayed in tables 2 and 3. A few noticeable properties are summarized below.

{\bf Result 1:} Let $\mathcal{G}_q^{\star}$ the dual Pauli graph of the definition 4, i.e. the graph attached to the projective line $\mathbb{P}_1(\mathbb{Z}_q)$, integral graphs of the following type are found

type 1: if $q=r$, with $r$ a prime, $\mbox{spec}(\mathcal{G}_q^\star)=\{(r+1)^1,(-1)^{r+1}\}$.

type 2: if $q=rs$, with $r,s$ two distinct primes, $\mbox{spec}(\mathcal{G}_q^\star)=\{q^1,1^q,(-r)^s,(-s)^r \}$.

type 3: if $q=rst$, with $r,s,t$ three distinct primes, $\mbox{spec}(\mathcal{G}_q^\star)=\{q^1,t^{rs},s^{rt},r^{st},(-1)^q,(-rs)^t,(-rt)^s,(-st)^r\}$.

type 4: if $q=4r$, with $r$ a odd prime, $\mbox{spec}(\mathcal{G}_q^\star)=\{q^1,2^{2r},0^{\psi(q)/2},(-4)^r,(-2r)^2\}$.

type 5: if $q=p^l$ is the power of a prime $p$, with $l>1$, $\mbox{spec}(\mathcal{G}_q^\star)=\{q^1,0^{\psi(q)-(p+1)},(-q/p)^p\}$.

type 6: in dimension $q \le 36$, cases not of the above type occur for $q= 18$ with $\mbox{spec}(\mathcal{G}_q^\star)=\{18^1,3^6,0^{24},(-6)^3,(-9)^2\}$, for $q= 24$ with $\mbox{spec}(\mathcal{G}_q^\star)=\{24^1,4^6,0^{36},(-8)^3,(-12)^2\}$ and for $q=36$ with $\mbox{spec}(\mathcal{G}_q^\star)=\{36^1,6^6,0^{60},(-12)^3,(-18)^2\}$.

{\bf Result 2:} Let $\mathcal{G}_q^{\star}$ the dual Pauli graph of the definition 4. Graphs and subgraphs of the following type are found in colum 6 of tables 2 and 3.

type 1: if $q=r$, with $r$ a prime, $\mathcal{G}_r^{\star}=K_{r+1}$.

type 2: if $q=rs$, with $r,s$ two distinct primes, copies of complete subgraphs of the form $p_j \times K_{p_i +1}$, $i,j \in\{1,2\}$, are found.

types 4, 5 and 6: For a dimension $q$ containing a square, the projective line $\mathbb{P}_1(\mathbb{Z}_q)$ contains bipartite or multipartite subgraphs.
 For $q \in \mathcal{A}_0$, they are of the type $K_{2,2}$, $K_{2,2,2}$, $\mbox{Cu}$ or $\mbox{BP}_6$ (see column 6 in Table 2). For $q \notin \mathcal{A}_0$, bipartite or multipartite subgraphs
 of the type $K_{4,4}$, $K_{4,4,4}$ and $\mbox{BP}_r$, $r \ne 6$ are found (see column 6 in Table 3). 

for type 3 and higher dimensions of the primorial type one gets result 3.

{\bf Result 3:} For primorial dimensions $q=N_r=\prod _{i=1}^r p_i$, $r>3$, the automorphism group of the graph $\mathcal{G}_q^*$ is a product of the symmetric groups $S_{p_i+1}$ arising from the factors of $q$. In addition, according to theorem 2, if Riemann hypothesis is satisfied, the size of the projective line follows as $\frac{|\mathbb{P}_1(\mathbb{Z}_q)|}{q \log \log q} > \frac{e^{\gamma}}{\zeta(2)} \sim 1.08$.

\subsection*{Comments }

Let us discuss shortly the results collected in tables 2 and 3. At first sight, the symmetries of a qudit in a dimension $q\in \mathcal{A}_0$ corresponding to a totally Goldbach number (in table 2) do not contrast with those at small dimensions $q \notin \mathcal{A}_0$, $q \le 36$ (in table 2). Let us restrict our reading of the tables to the cases where $q$ contains a square so that the total number $\sigma(q)$ of isotropic lines/maximal commuting sets, in (\ref{divisor1}), is strictly bigger than the size $\psi(q)$ of the projective line, in (\ref{proj1card}). Hence, our decomposition of the cardinality of the set of maximal commuting sets is as in column 3: $\sigma(q)=\psi(q)+(.)$ \cite{Planat2011}.

For these non square free cases, in table 2 (column 5) the two groups $G_{48}$ of the quartit and $G_{144}$ of the sextit are building blocks of the automorphism groups. Looking at the dual Pauli graph $\mathcal{G}_q^*$ as a simplicial complex, both groups correspond to the topology of the sphere and of the torus, respectively. Non trivial grid graphs or bipartite graphs are found as constituents of $\mathcal{G}_q^*$, as shown in column 6. In table 3, the graph  $\mathcal{G}_q^*$ for non square free dimensions shows symmetries of a different topology involving higher order symmetric groups and simpler bipartite subgraphs such as $\mbox{BP}_r$ (see the legend of the table for a definition). 

Dimension $24$ deserves a special attention because it is the only totally Goldbach number that violates the inequality of theorem 1. Note that dimension $24$ is also very special for lattices (such as the Leech lattice)
and marks a kind of phase transition for symmetries \cite{Sloane}, of a different type than the one described in \cite{PlanatSole2011}.

\section{On Goldbach pairs at primorial dimensions and RH}

 According to Hardy and Littlewood \cite{Hardy1922}, the number of Goldbach pairs $(p_1,p_2)$ of primes whose sum is the even integer $q$ is asymtotically given by the Hardy-Littlewood function $g(q)=R(q)\frac{q}{\log^2 q}$, with $R(q)=2 C_2 \prod_{p|q} \frac{p-1}{p-2}$, and $C_2=\prod_{p >2}(1-\frac{1}{(p-1)^2}) \sim 0.660$ is the twin prime constant. 

{\bf Proposition 1:} Let $N_r=2\cdots p_r$ be the primorial number of order $r$, we have $\lim_{r \rightarrow \infty} \frac{R(N_r)}{\log \log N_r}=e^{\gamma}$.

{\bf Proof: } One splits the infinite product for $C_2$ as

$$C_2=\prod_{i=2}^r (1-\frac{1}{(p_i-1)^2})\prod_{i=r+1}^{\infty}(1-\frac{1}{(p_i-1)^2}),$$

and one uses the equality $\frac{p_i-1}{p_i-2}=1/(1-\frac{1}{p_i-1})$ to obtain

\begin{equation}
R(N_r)=2\prod_{i=r+1}^{\infty} (1-\frac{1}{(p_i-1)^2})\prod_{i=2}^r (1+\frac{1}{p_i-1}).
\label{equation}
\end{equation}
Then, by the convergence of the infinite product for $C_2$

$$R(N_r)\sim 2 \prod_{i=2}^r (1+\frac{1}{p_i-1})=2\prod_{i=2}^r (1-\frac{1}{p_i})^{-1}=\prod_{i=1}^r (1-\frac{1}{p_i})^{-1}.$$

Finally, proposition 1 follows by using the same line of reasoning than in proposition 3 of \cite{PlanatSole2010}.
 
{\bf Conjecture 1:} The Hardy-Littlewood inequality $\frac{R(N_r)}{\log \log N_r}>e^{\gamma}$ holds for every $r>1$.

\begin{table}[ht]
\begin{center}
\small
\begin{tabular}{|r|r|r|r|}
\hline
$r$ & $N_r$ & $u_r-e^\gamma$   \\
\hline
$2$ & $6$ & $2.74$   \\
$10$ & $6.5 \times 10^9$ & $0.23$   \\
$10^2$ & $4.2 \times 10^{219}$ & $0.028$  \\
$10^3$ & $6.8 \times 10^{3392}$ & $0.0049$   \\
$10^4$ & $9.1 \times 10^{45336}$ & $0.0010$   \\
$10^5$ & $ 1.9 \times 10^{563920}$ & $0.00023$  \\
\hline
\end{tabular}
\label{tablecheck} 
\caption{Numerical evidence for conjecture 1 (in column 3).}
\end{center}
\end{table}

This proposed conjecture follows from numerical evidence as shown on table 4. In theorem 5, conjecture 1 is found to be equivalent to Riemann hypothesis.

{\bf Proposition 2:} For an even integer $q$, the following inequality holds
\begin{equation}
R(q)<\zeta(2)\frac{\psi(q)}{q}.
\label{2ineqs}
\end{equation}

{\bf Proof: } One introduces the ratio $x(q)=\frac{1}{\zeta(2)}\frac{R(q)}{\psi(q)/q}$ that reads.
$$x(q)=\frac{2 \prod_{p>2}(1-\frac{1}{(p-1)^2})}{\frac{4}{3}\prod_{p>2}\frac{1}{1-\frac{1}{p^2}}} \prod_{p>2,p|q}\frac{1}{\frac{3}{2}(1+\frac{1}{p})(1-\frac{1}{p-1})}.$$

By splitting the product as $\prod_{p\nmid q}()\times \prod_{p|q}()$ the ratio $x(q)$ is rewritten as

$$x(q)=\prod_{p>2,p\nmid q}\frac{1-\frac{1}{(p-1)^2}}{1-\frac{1}{p^2}}\prod_{p>2,p|q}(1+\frac{1}{p-1})(1-\frac{1}{p}).$$

The right hand side product $\prod_{p|q}()$ equals $1$ and the first one $\prod_{p>2,p\nmid q}$ is lower than $1$ so that Proposition 2 is satisfied.

{\bf Corollary 1:} The primorial numbers $N_r$ are champion numbers, i.e. left to right maxima of the function $q \rightarrow x(q)$. One gets $\lim_{r \rightarrow \infty} x(N_r)=1 $.

{\bf Proof: } According to \cite{Montgomery}, p. 90, or \cite{Sandor}, p. 238, the number of Goldbach pairs is such that there exists a constant $c_1$ with $R(q) < c_1 \frac{\psi(q)}{q}$. Using the same method than the one used in \cite{PlanatSole2010} for the champions of $\psi(q)/q$ one gets $c_1\equiv \zeta(2)$.

{\bf Theorem 5:} Conjecture 1 is equivalent to RH.

{\bf Proof:} If Riemann hypothesis is not true, then according to Theorem 4.2 of \cite{PlanatSole2010} and the inequality (\ref{2ineqs}) there are infinitely many $r$ such $\frac{R(N_r)}{\log\log N_r}<e^\gamma$, corresponding to as many violations of the Hardy-Littlewood inequality in Conjecture 1. 

Now we prove that Conjecture 1 follows from Riemann hypothesis.

It is known from theorem 3 (b) of \cite{Nicolas1983} that if RH is true then, for any $x \ge 2$,
$$f(x)=e^\gamma \log \theta(x) \prod_{p \le x}(1-\frac{1}{p})<1,$$
with $\theta(x)=\sum_{p\le x} \log p$ the first Tchebycheff function.

We introduce the new function

$$g(x)=\frac{e^\gamma}{2 C_2} \log \theta(x) \prod_{p \le x}\frac{p-2}{p-1}$$
and show below that, under RH, we have
\begin{equation}
g(x)<1.
\label{gf}
\end{equation}
Let us explicit the ratio
$$\frac{g(x)}{f(x)}=\frac{1}{C_2}\prod_{2<p\le x}\frac{p(p-2)}{(p-1)^2},$$
where $C_2=\prod_{2<p\le x} (1-\frac{1}{(p-1)^2})\prod_{p>x}(1-\frac{1}{(p-1)^2}).$
One immediately gets
\begin{equation}
\log g(x)=\log f(x)-\sum_{p>x}\log (1-\frac{1}{(p-1)^2}).
\label{loggf}
\end{equation}
Next, using the change of variables $y=x-1$, one is interested with the range of values such that $p-1>y$ so that $\log(1-\frac{1}{(p-1)^2})<\log(1-\frac{1}{y^2})$.

Replacing the sum by an integral at the right hand side of (\ref{loggf}), one obtains
$$-\sum_{p>x}\log(1-\frac{1}{(p-1)^2})<-\int_y^\infty \log(1-\frac{1}{u^2})du=-\left[u\log(1-\frac{1}{u^2})+\log(\frac{u+1}{u-1})\right]_y^{\infty} \sim \frac{1}{y}\sim \frac{1}{x},$$
at large $y$. By \cite[Theorem 3 (b)]{Nicolas1983} this is dominated by $\log f.$ Hence, by (\ref{loggf}) $\log g <0$ and the inequality (\ref{gf}) follows.

Then, by using the same reasoning that in \cite{Nicolas1983}, Conjecture 1 follows from the inequality $g(x)<1$ by observing that
$$g(p_r)=\frac{e^{\gamma}\log\log N_r}{R(N_r)}.$$
This completes the proof of Theorem 5.

\section{Discussion}

It is useful to encapsulate some of the results obtained for Goldbach pairs at small even dimensions and at primorial dimensions by introducing a measure that we denote the Goldbach defect.

{\bf Definition 6:} For a positive even integer $q$, one defines a Goldbach defect $\mbox{gd}_q$ from the relation $\mbox{gd}_q=\zeta(2)qR(q)-\psi(q)$.

{\bf Result 4:} The champions of the function $\mbox{gd}_q$ (the numbers at which $\mbox{gd}_q$ reaches a new record) form the sequence

$\mathcal{C}=\{ 2, 4, 6, 12, 18, 24, 30, 42, 54, 60, 84, 90, 120, 150, 180, 210, 270,300,330,390, 420,\\ 510, 570, 630, 780, 840, 990, 1050, 1260, 1470, 1650, 1680, 1890,  2100, 2310,2730, 3150,3360,\\ 3570, 3990, 4290, 4410, 4620,    5250, 5460, 6090, 6510, 6930, 7770, 7980, 8190,
9030, 9240 \ldots\}$
  
The sequence $\mathcal{C}$ is found to be a subset of the champions of the cototient function (A051953 in Sloane's encyclopedia). By its definition, sequence $\mathcal{C}$ encompasses the even totally Goldbach numbers, except for the number $10$ (see definition 1)
and the primorial numbers (see Conjecture 1 and Corollary 1). None of the (small) even dimensions in table 3 belongs to the sequence. INumbers in $\mathcal{C}$ satisfy 
$$\frac{\zeta(2)\psi(i)}{i \log\log i}>\frac{R(i)}{\log\log i}>e^{\gamma} ~\mbox{for}~\mbox{all}~i \in \mathcal{C},~ i \ne 2.$$

In future work, the Goldbach defect may turn to be useful for discussing the change in the symmetries of the Pauli graphs. The first not square free and not totally Golbach dimension in the sequence $\mathcal{C}$ is $q=60$. The cardinality of maximal commuting sets for this case decomposes as $\sigma(60)=\psi(60)+(.)$, i.e. $168=144+24$.

In conclusion, we found a cornerstone between the arithmetic of qudits, the Goldbach pairs and Riemann hypothesis. It may be that the loss of coherence of a classical system, in comparison to a quantum system, crucially depends on its high dimensionality. The inequalities of theorem 2 and 4 would serve to define an entropy. Mixtures of multiple qudit structures also have an attractive nested arithmetical and geometrical structure that is worthwhile to investigate in detail (see Table 1 in \cite{Planat2011} and \cite{SanPlanat2001}).

\end{document}